\documentclass[fleqn,usenatbib]{mnras}

\newcommand{\nai}{Na\,I}

\newcommand{\galex}{\emph{Galex}}
\newcommand{\tess}{\emph{TESS}}

\newcommand{\gaia}{\emph{Gaia}}
\newcommand{\ktwo}{\emph{K2}}

\newcommand{\kepler}{\emph{Kepler}}

\newcommand{\kms}{\mbox{km\,s$^{-1}$}}

\newcommand{\msun}{M$_{\odot}~$}
\newcommand{\msune}{M$_{\odot}$}

\newcommand{\lsun}{L$_{\odot}~$}
\newcommand{\lsune}{L$_{\odot}$}
\newcommand{\rsune}{R$_{\odot}$}

\newcommand{\rearth}{$R_{\oplus}$}

\newcommand{\teff}{\ensuremath{T_{\rm eff}}}

\newcommand{\kicstar}{KIC~12557438}
\newcommand{\epicstar}{EPIC~201637175}

\usepackage{newtxtext,newtxmath}
\usepackage[T1]{fontenc}
\usepackage{ae,aecompl}

\usepackage{graphicx}	% Including figure files
\usepackage{amsmath}	% Advanced maths commands
\usepackage{amssymb}	% Extra maths symbols

\title[]{Monitoring of the D Doublet of Neutral Sodium during Transits of Two ``Evaporating" Planets}

\author[Gaidos, Hirano, \& Ansdell]{
Gaidos, E.,$^{1}$\thanks{E-mail: gaidos@hawaii.edu}
T. Hirano,$^{2,3}$
M. Ansdell,$^{3,4,5}$
\\
$^{1}$Department of Earth Sciences, University of Hawai'i at M\={a}noa, Honolulu, HI  96822, USA\\
$^{2}$Department of Earth \& Planetary Sciences, Tokyo Institute of Technology, Tokyo 152-8551, Japan\\
$^{3}$Institute for Astronomy, University of Hawaii at M\={a}noa, Honolulu, HI 96822, USA\\
$^{4}$Center for Integrative Planetary Science, University of California at Berkeley, Berkeley, CA 94720, USA\\
$^{5}$Department of Astronomy, University of California at Berkeley, Berkeley, CA 94720, USA\\
}

\date{Accepted to MNRAS, 24 Feb. 2019}

\pubyear{2019}

\begin{document}
\label{firstpage}
\pagerange{\pageref{firstpage}--\pageref{lastpage}}
\maketitle

\begin{abstract}
Spectroscopic transit detection of constituents in winds from ``evaporating" planets on close-in transiting orbits could provide desperately needed information on the composition, formation, and orbital evolution of such objects.  We obtained high-resolution optical spectra of the host stars during a single transit of Kepler-1520b and two transits of K2-22b to search for transient, Doppler-shifted absorption in the D lines of neutral sodium.  Sodium should be released in the same silicate vapor wind that lofts the dust responsible for the periodic ``dips" in the light curve.  We do not detect any absorption lines with depths $>$30\% at the predicted Doppler-shifted wavelengths during any of the transits.   Detection sensitivity is limited by instrumental resolution that dilutes the saturated lines, and blurring of the lines by Doppler acceleration due to the short orbital period of the planet and long integration times for these faint stars.  A model of neutral sodium production, escape, and ionization by UV radiation suggests that clouds of partially ionized sodium that are comparable in size to the host stars and optically thick in the D lines could accompany the planets.  We consider the prospects for future detections brought about by the \tess\ all-sky survey of brighter stars and the advent of high-resolution spectrographs on Extremely Large Telescopes.
\end{abstract}


%%%%%%%%%%%%%%%%%%%%%%%%%%%%%%%%%%%%%%%%%%%%%%%%%%

%%%%%%%%%%%%%%%%% BODY OF PAPER %%%%%%%%%%%%%%%%%%

\section{Introduction}

Transiting planets are distinguished from other variable stars by their Keplerian periodicity, the distinctive shape of the transit light curves, and the constancy of those light curves from transit to transit.  \citet{Rappaport2012} identified an unusual system, \kicstar\ (now Kepler-1520) in \kepler\ mission data where the transits occur with precise regularity ($P = 15.7$~hr), but the depth and shape of the transits vary with no discernible pattern, and some expected transits were not detected.  The current model to explain this phenomenon invokes an airless, Mercury-size planet (designated ``b"), too small to be detected itself, on a close-in orbit where it is heated by incident stellar irradiation to $>$2000~K and is evaporating \citep{Rappaport2012,Brogi2012,Perez-Becker2013,vanWerkhoven2014}.  The resulting rock vapor wind lofts dust particles that periodically sweep across our line of sight, partially obscuring the star.  The asymmetry of the transit lightcurve means that the cloud subtends a significant fraction of the stellar disk.  A second system with significantly shallower events (KOI-2700b, 21.8~hr) was also identified among \kepler\ target stars \citep{Rappaport2014}.   A reaction wheel failure ended observations of the \kepler\ field in May 2013, but the reincarnation of \kepler\ in two-wheel mode as the \ktwo\ mission led to the discovery of a similar system, \epicstar{} or K2-22 with a ``b" object on a 9.14~hr orbit around an M dwarf star \citep{Sanchis-Ojeda2015}.

\kepler\ and \ktwo\ observations are in a single, wide band-pass.  An airless planet will produce a transit that is the same depth at all wavelengths, but scattering by small grains and absorption by gases in a wind or escaping atmosphere should impart a wavelength-dependent signal. \citet{Croll2014} found no difference between transits of Kepler-1520b in the infrared ($K$-band, 2.2\micron) and at visible wavelengths, indicating that the particles must be larger than $\sim$1\micron.  On the other hand, \citet{Bochinski2015} reported that the depth of a single transit of Kepler-1520b simultaneously observed in different optical pass-bands varied by about 25\%, indicating particle sizes of 0.25-1$\mu$m.  \citet{Colon2018} found no evidence for wavelength dependent variation in transits of K2-22b, albeit with limited data having limited photometric precision.

Presuming that these planets never possessed or lost any atmospheres of light, volatile elements, the wind posited to loft these grains must derive from an evaporating crust, and thus should include moderately volatile elements such as the alkali metals (Na, K, etc.).  These elements could also be released as the grains evaporate once exposed to the full irradiance by the central star.  Sodium has a 50\% condensation temperature of $\approx$1000K and neutral sodium (Na I) might be detected in absorption by its strong "D" doublet at 5889.95 and 5895.92 \AA.  The planet Mercury has an exosphere that contains photon-desorbed sodium which has been detected from the ground \citep{Potter2013} and by the \emph{Messenger} spacecraft \citep{Cassidy2015}.  Neutral sodium has been detected in the atmospheres of some giant transiting exoplanets \citep{Charbonneau2002,Redfield2008,Zhou2012,Sing2012,Burton2015}.  \citet{Ridden-Harper2016} conducted an inconclusive search for Na I in any atmosphere of 55 Cancri e, a 2.2\rearth\ planet on a 17.7~hr orbit around a K0-type dwarf.  

Neutral sodium produced by a transiting, evaporating planet should manifest itself as narrow lines superposed on the pressure-broadened lines of the stellar photosphere, as well as any emission from the stellar chromosphere.  The apparent radial velocity shift (relative to the star) of gas that is co-moving with a planet on a transiting orbit will vary during the transit between $\pm 2\pi R_* \sqrt{1-b^2}/P$, where $R_*$ is the stellar radius, $b$ the transit impact parameter, and $P$ the orbital period.  This will be as large as $\pm55$~\kms\ for Kepler-1520b, and $\pm90$ \kms\ for K2-22b (see Sec. \ref{sec:hosts} for the parameters used for these calculations).  This line may also be joined by a fixed line due to Na I in the intervening interstellar medium that is Doppler-shifted relative to the stellar lines by the star's peculiar velocity with respect to the Local Standard of Rest.  

For a thermally broadened line from \nai\ atoms at temperature $T$ (pressure broadening and the intrinsic line width are negligible), the width (FWHM) of the line in a short-exposure spectrum is
\begin{equation}
\label{eqn:linewidth}
    \frac{\Delta \lambda}{\lambda} = \sqrt{\frac{8 (\ln 2) k_B T}{\mu c^2}},
\end{equation}
where $k_B$ is Boltzmann's constant, $\mu$ is the atomic weight of Na atoms, and $c$ is the speed of light.  For $T=1000$K, $\Delta \lambda \approx 27$~m\AA, and the line will be unresolved in any spectrum with resolution $R < 2 \times 10^5$.  At the start of or even prior to the transit, absorption by an accompanying wind or cloud that is much smaller than the stellar disk would appear as a blue-shifted line.  This line will shift to the red over the course of the transit ($\sim$1~hr) and then eventually disappear.  If the integration time $\delta t$ is a substantial fraction of the transit duration $T$, then the line will also be broadened by orbital acceleration by an amount 
\begin{equation}
\label{eqn:blurring}
    \frac{\Delta \lambda}{\lambda} = \frac{2 \pi R_* \sqrt{1-b^2}T}{c P},
\end{equation}
or about 0.4\AA\ and 0.8\AA\ in 15-minute spectra of Kepler-1520 and K2-22, respectively.  Alternatively, a torus-like cloud that occupied the entire orbit of the planet would produce lines as broad as 2.1\AA\ and 3.5\AA\ respectively.    

The Na gas co-produced with the dust responsible for the transits of Kepler-1520b and K2-22b should be optically thick in the D lines.  Neglecting increased path length due to scattering by the dust particles that produce the transits, the optical depth in the line centres  is $\tau = \Sigma \sigma / \Delta \lambda$ where $\Sigma$ is the column density of \nai\ atoms, $\sigma$ is the line cross-section, and $\Delta \lambda$ is the width of the line.   The optical depth is:
\begin{equation}
\label{eqn:opticaldepth}
\tau \approx \frac{\Sigma \sigma}{\lambda} \sqrt{\frac{\mu c^2}{k_B T}}.
\end{equation}
For the D lines, $\sigma = 9.8 \times 10^{-14}$~cm$^2$~\AA\ and thermally broadened lines (see above) will  become saturated by $\Sigma \sim 10^{12}$~cm$^{-2}$.  We can relate $\Sigma$ to the depth of the transit $\delta$ if we assume if the Na is produced by evaporation of a uniform population of grains of diameter $d$ and density $\rho$ that produce the transit.  Then:
\begin{equation}
\label{eqn:coldensity}
\Sigma = \frac{2 \rho d \delta f_{\rm Na}}{3 \mu}
\end{equation}
where $f_{\rm Na}$ is the mass fraction of Na that remains volatilized.   (A more realistic calculation, performed in Sec. \ref{sec:expectations}, must also account for photoionization by UV photons from the star as well as recombination.)  Combining Eqns. \ref{eqn:opticaldepth} and \ref{eqn:coldensity},
\begin{equation}
\label{eqn:opticaldepth2}
\tau \approx \frac{2 \sigma \rho d \delta c f_{\rm Na}}{3 \lambda \sqrt{\mu k_B T}}.
\end{equation}
 For $\rho = 2$~g~cm$^3$, $d = 1$\micron\, $f_{\rm Na} = 0.01$, and $T = 1000$K, the optical depth in the D lines would be $3 \times 10^5\delta$, or $\sim 10^3$ for these objects.  Because the lines are saturated the equivalent width will be of order the thermally broadened line width, but will nevertheless slowly increase with column density as the wings of lines contribute.  If the cloud occults a fraction of the stellar disk then the equivalent width will be proportionally less.   And because the lines are unresolved the observed line depth will be the ratio of the equivalent width to either the spectral resolution or -- if greater -- the Doppler blurring due to motion of the gas during the integration.      
   
Detection of alkali metals would be compelling evidence for the evaporating planet model because the short lifetime of \nai\ against photoionization due to the proximity to the star demands a replenishing source (see Sec. \ref{sec:expectations}).  Their detection would also spur follow-up observations to probe the spatial and velocity structure of the gas cloud and the composition of a dying exoplanet. Nondetections mean that the source of the wind and dust is devoid of these elements, e.g. a planet has lost its silicate mantle and an exposed iron core is evaporating \citep{Perez-Becker2013}, or that there is a problem with the evaporating planet model itself.  These observations are also a potential test of explanations for the formation of close-in planets.  Formation from warm material close to the star \citep{Chiang2013} should yield  planets depleted in volatiles such as Na, compared to planets that formed further out in the disk and subsequently inward migrated.

However, the short transit times, the narrowness of the hypothetical lines, and faintness of these stars ($V=16.7$ for Kepler-1520 and $V=15.6$ for K2-22) requires rapid cadence combined with high spectral resolution at high sensitivity, meaning that 8-10 m telescopes \emph{must} be employed.   Thus we observed transits of both Kepler-1520b and K2-22b with the High Dispersion Spectrograph (HDS) on the 8.2 m Subaru telescope on Maunakea.  We revisit the properties of the host stars in Section \ref{sec:hosts}, present the observations and data reduction in Section \ref{sec:observations}, and our analysis in Section \ref{sec:results}. We compare these results to our expectations for \nai\ around these two objects in Section \ref{sec:expectations} and in Section \ref{sec:discussion} we summarize and explore the potential for the NASA \tess\ mission \citep{Ricker2014} to discover more such systems around brighter, more amenable stars, and the promise of upcoming generation of Extremely Large Telescopes to make more sensitive observations.

\section{Revised Properties of the Host Stars}
\label{sec:hosts}

{\bf Kepler-1520/KIC~12557548:} The light curve of Kepler-1520 has been thoroughly examined \citep[see, e.g.,][and references therein]{Schlawin2018}.  The \gaia\ DR2 parallax of the star \citep{Gaia2018} is $1.62 \pm 0.03$ mas.  The Bayesian inversion of \citet{Bailer-Jones2018} places the star at a distance of $608\pm11$~pc.  The three-dimensional reddening map of \citet{Green2018} gives $E(B-V)=0.065$ which we converted to $A(K_s) = 0.02$ using the coefficients of \citet{Yuan2013}.  These yield an absolute $K$ magnitude of $M_K=4.38\pm0.04$, and a mass of $0.70 \pm 0.03$\msun \citep{Mann2018}.  This is consistent with a K4.5/5 spectral type on the main sequence, in agreement with the analysis of \citet{Rappaport2012}.  Adopting a $K$-band bolometric correction of 2.19 magnitudes \citep{Pecaut2013}, an effective temperature of $4440 \pm 70$ K based on other HDS data \citep{Schlawin2018} and applying the Stefan-Boltzmann law, we arrive at a luminosity of 0.19\lsun and a radius of $0.73 \pm 0.02$\rsune.  For the planet's orbital period of 15.685\,hr, we estimate the transit duration of a much smaller object (at impact parameter $b=0$) to be 1.3 hours.  Convolving a simulated transit signal using the limb-darkening parameters from \citet{Claret2004} with the 30-min cadence of \kepler\ produces a transit signal lasting 1.7\,hr, consistent with the observations \citep{Rappaport2012}, and indicating that the occulting dust cloud is smaller than the star.  

\citet{Croll2014} determined the barycentric radial velocity of Kepler-1520 to be -36.3~\kms\ which, combined with the \gaia\ DR2 parallax and proper motions, yields a space motion of $(U,V,W)=(-34.8,-34.1,+1.6)$ \kms\ (barycentric) or $(-45.9,-46.1,-5.7)$ \kms\ with respect to the Local Standard of Rest (LSR) of \citet{Schonrich2010}.  This motion is not consistent with any of the known young co-moving groups cataloged by {\tt Banyan}~$\Sigma$ \citep{Gagne2016} but is consistent with the older ``thick disk" population of stars \citep{Fuhrmann2004}.  
% Gaia values: PM 0.321+/-0.065, 11.146 +/-0.055, plx=1.617 +/- 0.03

\begin{figure}
	\includegraphics[width=\columnwidth]{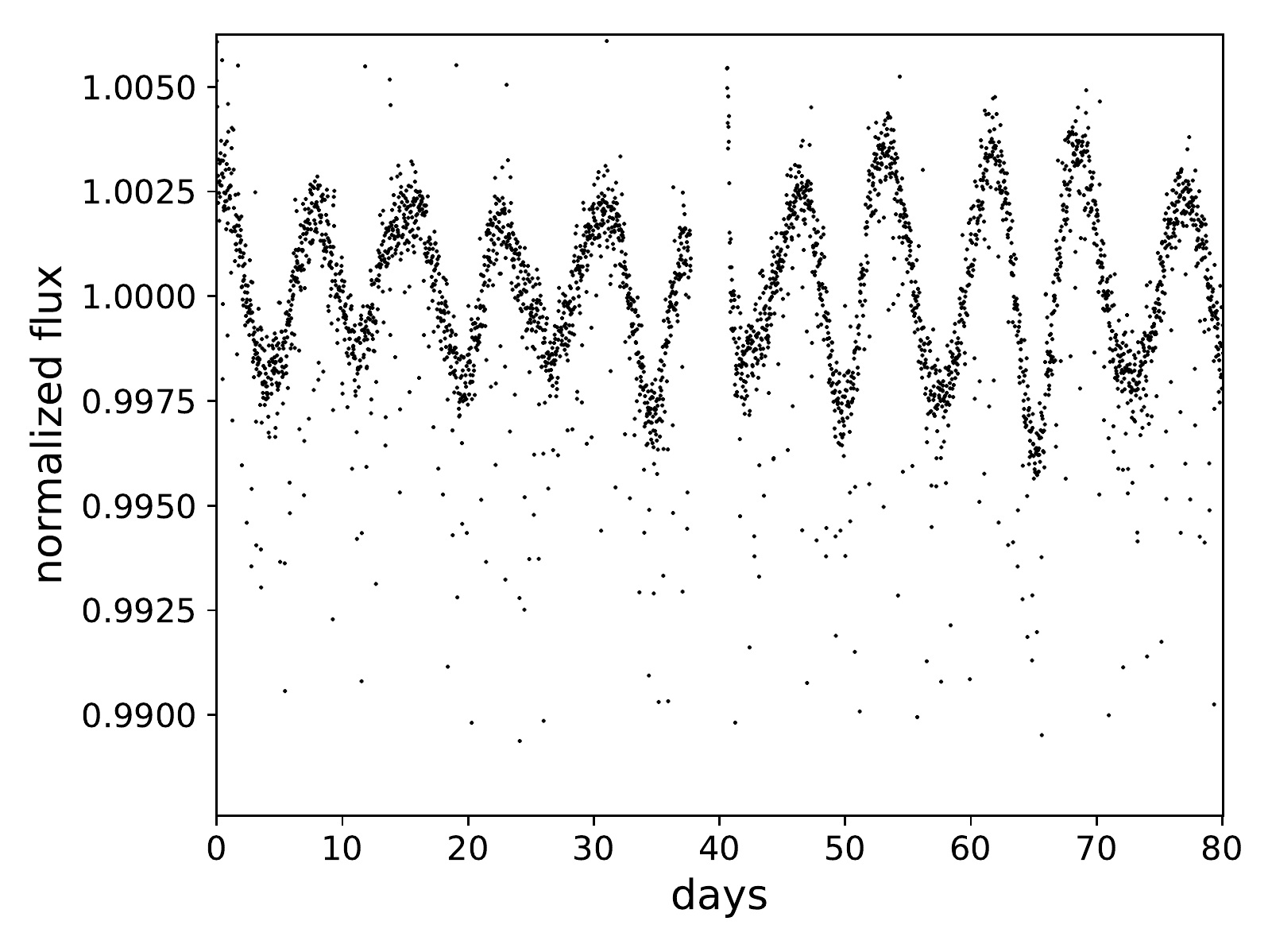}
    \caption{De-trended \ktwo\ light curve of K2-22 (\epicstar), showing the 7.7 day periodicity that may be (half of) the rotational period.  The transit events from the disintegrating planet can be seen. The break at 40 days was the result of an interruption in spacecraft pointing and observations.}  
    \label{fig:k2-22-lightcurve}
\end{figure}

\begin{figure}
	\includegraphics[width=\columnwidth]{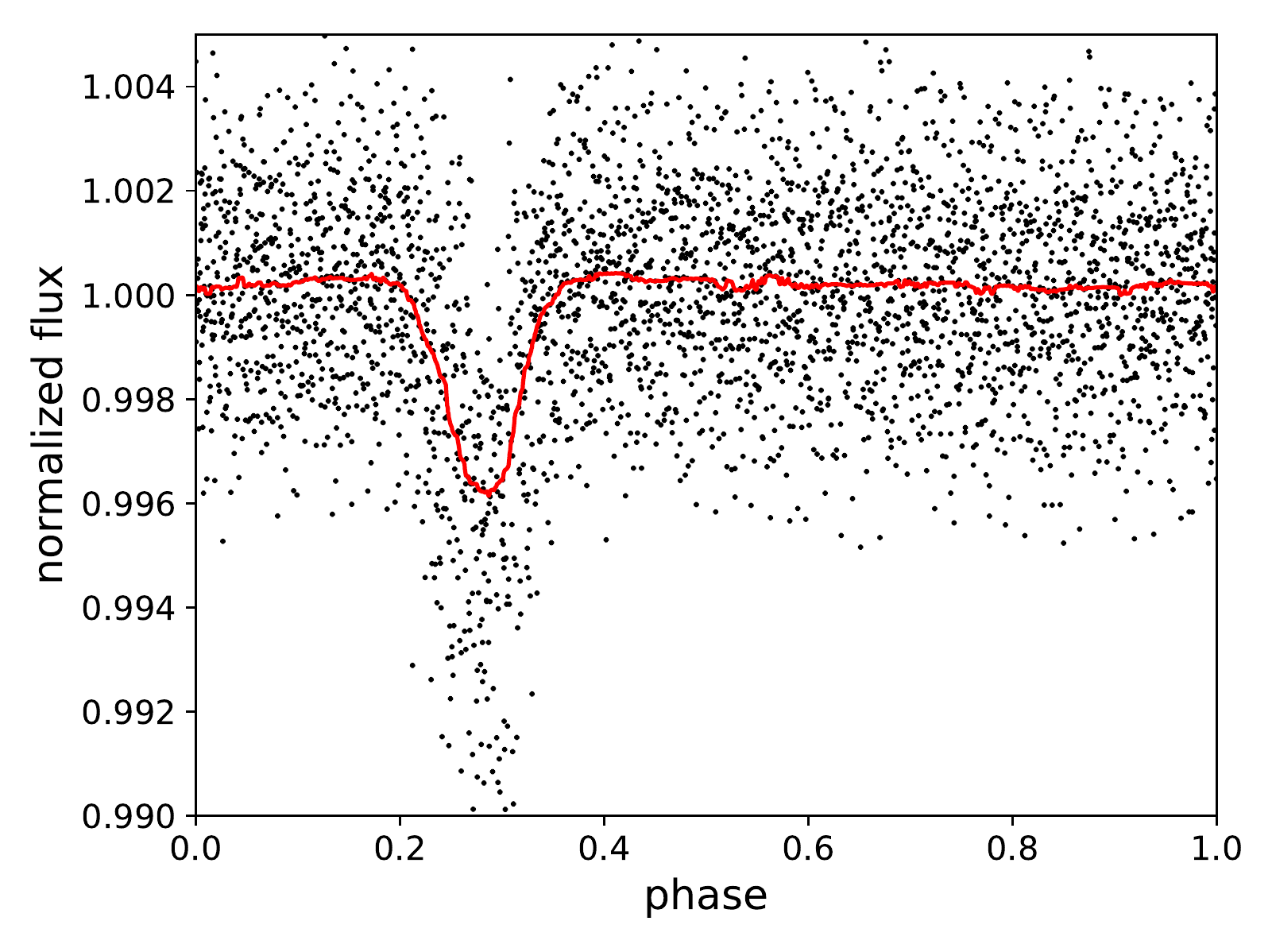}
    \caption{De-trended light curve of K2-22, phased to the disintegrating planet signal at 9.14 hours.  The line is a running median.}  
    \label{fig:k2-22-phased}
\end{figure}

{\bf K2-22/EPIC 201637175:} The \emph{K2} lightcurve of K2-22 (Fig. \ref{fig:k2-22-lightcurve}) shows the 1.4\,hr-long transit-like signal with a period of 0.38 days (Fig. \ref{fig:k2-22-phased}).  A periodic signal of 7.68 days is also recovered in the Lomb-Scargle analysis and we assume this is either the rotation period of the star or half that value. \citet{Sanchis-Ojeda2015} performed an autocorrelation analysis and found that the peak at twice the period is larger.  Our analysis finds that the first maximum (at a period of 7.3 days) is slightly larger than the second peak, but this is possibly an effect of detrending.  A 15-day rotation period would be consistent with the older age of a field M dwarf:  \citet{Douglas2016} found that \emph{single}, 650 Myr-old Hyades stars with similar masses (0.6\msune) have rotation periods of 10-15 days but \emph{close binary} stars could have significantly shorter periods consistent with a 7.68-day period.  K2-22 is a binary but with a separation wider than those which could affect rotational history (see below).  Additional time-series photometry will be required to definitively resolve this question.

Based on the \gaia\ DR2 parallax of $4.07\pm 0.05$~mas and a corresponding distance of $244 \pm 3$~pc \citep{Bailer-Jones2018}, a \citet{Green2018} reddening of $E(B-V) =0.04$ mag and hence $A_{K_s} = 0.01$, the absolute $K$-magnitude of K2-22 is $M_K = 4.98\pm 0.03$, which translates to a main sequence mass of $0.60 \pm 0.03$\msun \citep{Mann2018}.  Using a spectroscopic \teff\ of $3780 \pm 90$K \citep{Sanchis-Ojeda2015} the $K$-band bolometric correction is 2.48, leading to a luminosity of 0.08\lsune.  Combining this with the \teff\ and Stefan-Boltzmann law the radius is $0.68 \pm 0.03$\rsune.  The maximum transit duration for a ``small" planet on a 9.15-hr orbit around such a star is 1.05~hr.  Simulating the lightcurve of a transit with the parameters given above, the limb darkening parameters of \citet{Claret2004} and a 30-min cadence produces an event with a duration of 1.45~hr.  This is also consistent with the observations and means that the body or cloud responsible for the obscuration must have an extent that is smaller than the central star large enough to produce an asymmetric transit shape and certainly no smaller than 4.5\rearth, the minimum size required to produce the median transit depth if the cloud is optically thick.

Using our HDS spectra (see Sec. \ref{sec:observations}) we estimate a barycentric radial velocity of -9.2 km~sec$^{-1}$.  Combining this with the \gaia\ DR2 parallax and proper motion, we arrive at a space motion of $(U,V,W)=(-23.6, -10.3, -21.4)$ \kms\ with respect to the Sun or (-34.7,-22.5,-28.7) with respect to the \citet{Schonrich2010} LSR.  K2-22 is not a member of any of the well-known nearby clusters and co-moving groups \citep{Gagne2018}.  Like Kepler-1520, its space motion places it well within the zone of the ``thick disk" population as defined by \citet{Fuhrmann2004}.  \citet{Sanchis-Ojeda2015} identified a putative 2" companion by a $z$-band imaging at the Subaru telescope.  This star also appears in the \gaia\ DR2 catalog and it is definitely a physical companion: its parallax and proper motion are within 1$\sigma$ of those of the primary.  Using the more precise parallax of the primary, its $M_K \approx 7.0$, its mass is about 0.29\msun\ (a mid-type M dwarf) and the projected separation is 470\,AU.   

\begin{figure}
	\includegraphics[width=\columnwidth]{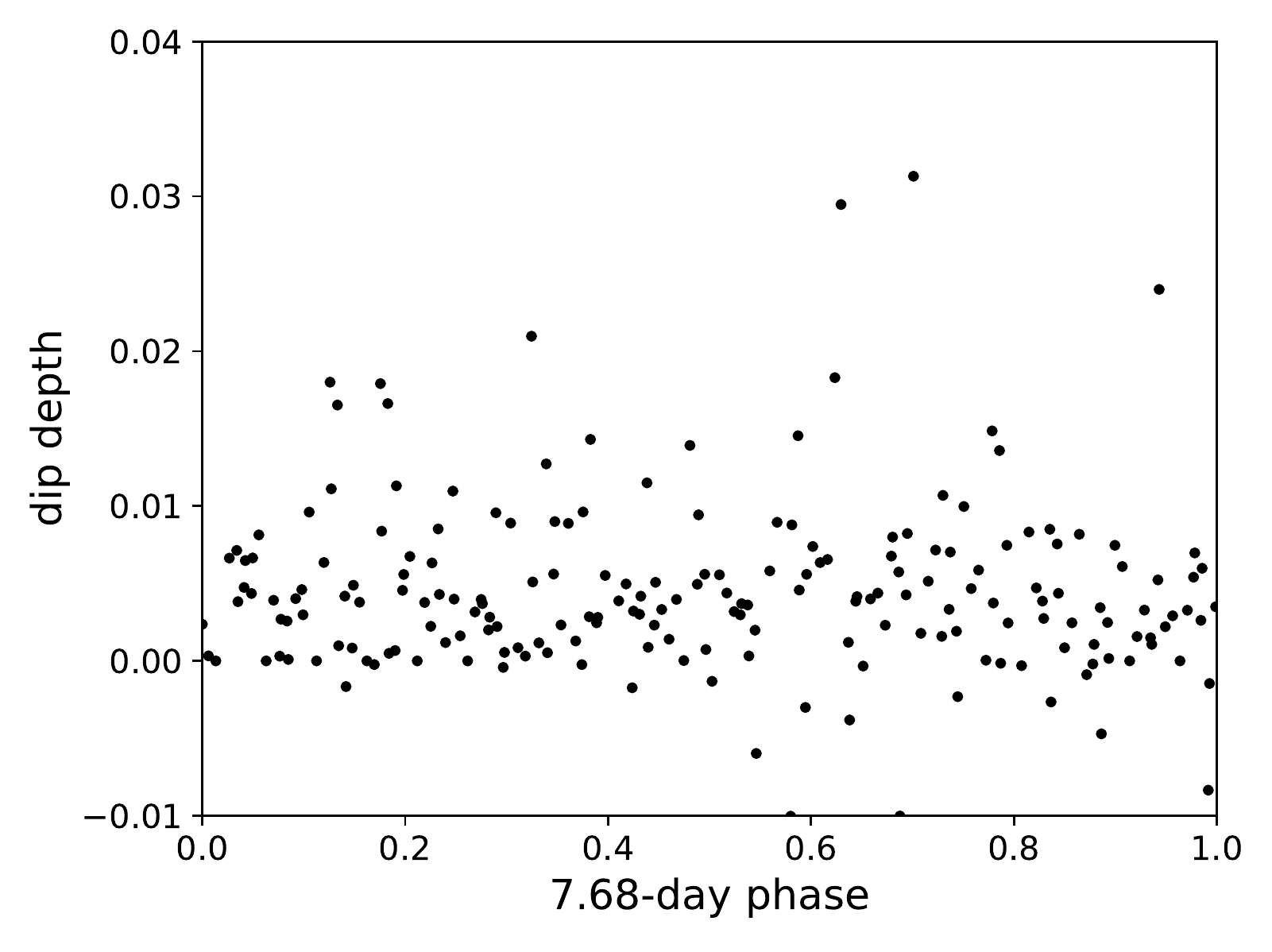}
    \caption{Transit depth vs. phase of the 7.68 day signal in the \kepler\ light curve of K2-22, which may correspond to the stellar rotation period (or half its value).}  
    \label{fig:k2-22-depth-rotation}
\end{figure}

Figure \ref{fig:k2-22-depth-rotation} plots the depth of individual transits vs. the phase of the 7.68 day periodic (rotational) signal.  No trend of transit depth with phase is seen, contrary to expectations if evaporation is sensitive to ultraviolet emission or charged particles emanating from major spot groups that also produce the rotational variability.  On the other hand, \citet{Doyle2018} report no correlation between flares on M dwarfs and rotational phase, possibly because the activity is primarily associated with polar spots. 

\section{Observations and Data Reduction}
\label{sec:observations}

Spectra of Kepler-1520 and K2-22 were obtained with the High Dispersion Spectrograph (HDS) on the Subaru telescope \citep{Noguchi2002} with the standard ``Rb" setup, which covers $5337-7994$ \AA, and a 0.8 arcsec slit which delivers $R=45,000$.  Individual integration times were 900 sec for both science targets.  A single transit of Kepler-1520b was observed on the night of UT 2014 August 13 (23 integrations), and two transits of Kepler-1520b were observed on UT 2016 January 26 and 29 (26 and 22 integrations, respectively).  Observations of spectrophotometric calibrator stars were performed prior to and subsequent to the transit observations:  BD+33 2642 (an O-type post-AGB star) and Feige 110 (an O-type subdwarf) in the case of Kepler-1520, and HD 93521 (an O-type star) in the case of K2-22.  Flat field and Th-Ar arc wavelength calibration spectra were obtained both at the beginning and end of each night.  For the Kepler-1520b transit, the Moon was 95\% illuminated and was 70 deg away. However for the first K2-22b transit observation the 91\% illuminated Moon was only $6-8$ deg away and for the second observation the 69\% Moon was $27-29$ deg away and the sky background was significant.  Clouds were present for the Kepler-1520 observations but both sets of of K2-22 observations occurred during photometric conditions.

We reduced the HDS data using {\tt IRAF}, performing a non-linearity correction \citep{Tajitsu2010}, bias subtraction, flat fielding, and scattered-light subtraction before extracting one-dimensional spectra. The red parts of HDS spectral images ($>6500$ \AA) are known to suffer from fringing.  For that spectral region, we used the IRAF task {\tt apnormalize} to create a normalized master flat frame, by which individual science frames were divided for flat fielding.  Since the HDS data, especially for K2-22 taken on January 26, contained high sky background levels (from the Moon), we implemented a careful extraction of the spectra; we first inspected the two-dimensional echelle spectra visually and checked the level of sky background along the spacial direction. As a result, we found that the high-level sky background extended up to $\pm 7$ pixels from the peak (centre) pixels of stellar spectra and that stellar light covered approximately $\pm 3$ pixels from the peak pixels.  Summing up the counts between $-6$ to $-3$ pixels and between $3$ to $6$ pixels in the spatial direction relative to the stellar aperture pixels, we extracted the spectra of the sky background.  These background spectra were then subtracted from the star plus sky spectra that were obtained by summing up $\pm 3$ pixels.  Errors were calculated as read noise plus electron counting noise from all included pixels added in quadrature.  Wavelengths were calibrated by identifying many emission lines in comparison Th-Ar lamp spectra taken before and after the science observations.  Due to the fringing at longer wavelengths, our primary analysis is based on data obtained in the blue region/CCD.

Following \citet{Hirano2018}, we estimated the barycentric radial velocity  for K2-22. We cross-correlated the reduced HDS spectra against the numerical binary mask \citep[M2 mask;][]{Bonfils2013} for each echelle order and inspected the peak of the cross-correlation function to estimate the absolute radial velocity. To take into account the possible instrumental drift of the HDS spectrograph during a night, we also cross-correlated the spectral segments including telluric absorption lines against the theoretical telluric transmittance of the atmosphere over Maunakea, created by using line-by-line radiative transfer model \citep{Clough2005}. After subtracting thus measured instrumental drifts (typically less than a few hundred m sec$^{-1}$) and applying the barycentric correction, we found a barycentric radial velocity of $-9.2 \pm 0.2$ km sec$^{-1}$ for K2-22. Note that the velocity error is computed based on the scatter of velocity measurements between individual frames.

\section{Results and Analysis}
\label{sec:results}

\begin{figure}
	\includegraphics[width=\columnwidth]{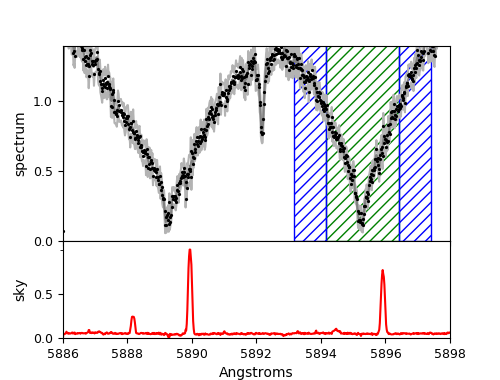}
    \caption{Spectrum of Kepler-1520 in the vicinity of the Na I D lines obtained on UT 13 Aug 2014.  The D1 line core and wing regions used to look for time-variable absorption are plotted as green and blue hatched zones, respectively.  The upper and lower grey lines mark $\pm 2\sigma$ errors. The bottom red curve is the sky background.}
    \label{fig:kepler-1520_sumspec}
\end{figure}

The summed spectrum in the vicinity of the \nai\ D lines of Kepler-1520 obtained on UT 2014 August 13 is plotted as points in Fig. \ref{fig:kepler-1520_sumspec}, along with a  second-order Savitzky-Golay filtered version  \citep[solid line,][]{Savitsky1964}.  In Savitsky-Golay filtering, low-order polynomials are fit to successive subsets of the data.  This smoothing method has been widely used to resolve overlapping peaks in chemical spectra and is thus useful in identifying multiple absorption features.  The only feature besides the Doppler-shifted stellar photosphere lines (-36.9 \kms\ or about 0.7\AA\ including barycentric correction) are artefacts produced by subtraction of the background sky lines (red curve at bottom of plot).  Besides the prominent telluric \nai\ lines, the weaker sky lines near 5888.2 and 5894.5\AA\ are from OH airglow \citep{Osterbrock1997}.  In the direction of Kepler-1520, the Sun's motion with respect to the LSR -- and the likely rest frame of any interstellar clouds -- is -15 \kms, based on a solar motion of $(UVW)$ = (11.1,12.2,7.3) \kms\ and solar apex of ($\alpha,\delta) =(267,+23)$ deg \citep{Schonrich2010}.  Since the barycentric correction from Maunakea at the epoch of observation is -0.6 \kms, we expect any ISM \nai\ absorption to be red-shifted with respect to the stellar lines by 0.43\AA.  A detailed inspection of the spectrum shows no significant feature at that location, nor at the appropriate wavelength in spectra of K2-22 (see below).

\begin{figure}
	\includegraphics[width=\columnwidth]{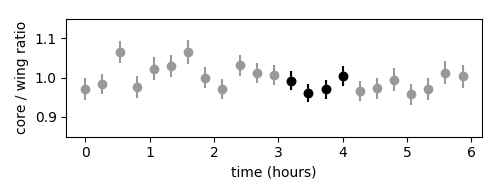}
    \caption{Ratio of the flux in the core of the D1 (red) line of Na I relative to that in the wings of the spectrum of Kepler-1520 (see Fig. \ref{fig:kepler-1520_sumspec}) before, during (black points), and after the transit of the evaporating planet on UT 13 August 2014.}
    \label{fig:kepler-1520_time}
\end{figure}

To search for changes in the line intensity in the core that could indicate the transient presence of circumstellar neutral Na, we integrated the intensity over a wavelength ranges corresponding to the Doppler shift of the transiting planets assuming $b=0$ (2.2\AA\ for Kepler-1520 and 3.5\AA\ for K2-22)  and normalizing by the flux in two 1\AA-wide regions immediately to the blue and red of the central region (green and blue hatched regions respectively in Fig. \ref{fig:kepler-1520_sumspec}).  The D1 line is used since the D2 line is more affected by telluric H$_2$O lines \citep{Lundstrom1991,Lallement1993}.  Fig. \ref{fig:kepler-1520_time} shows the the time series of this flux ratio.  No appreciable change in absorption in the core of the D1 line is seen during the transit.

\begin{figure}
	\includegraphics[width=\columnwidth]{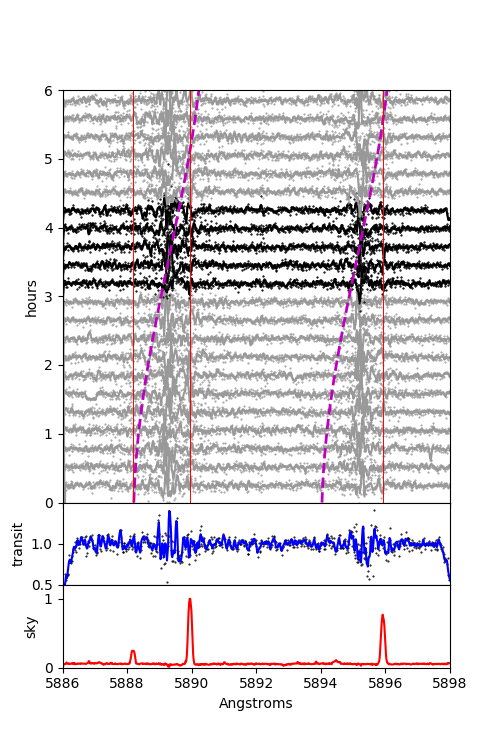}
    \caption{Individual, normalized spectra of \kepler-1520 in the vicinity of the Na I D doublet during a transit of \kepler-1520b on UT 13 August 2014.  The constant stellar spectrum has been divided out.  Both individual measurements (points) and a Savgol-filtered curve (solid lines) are plotted.  Grey spectra were obtained outside of the predicted transit, while black spectra were obtained within the transit.  The dashed magenta lines show the expected Doppler shift of the disintegrating planet and any co-moving sodium cloud.  The middle panel contains the sum of the spectra taken during transit and shifted to the rest frame of the planet.  The bottom red plot is the spectrum of the sky background and red vertical lines mark telluric line emission.}
    \label{fig:kepler-1520_series}
\end{figure}

The individual spectra are normalized by the mean stellar spectrum and plotted (along with Savitzky-Golay filtered versions) in Fig. \ref{fig:kepler-1520_series}.  Black points/curves are those observations obtained during the transit event, after correcting for effect of the finite cadence of \kepler{}/K2 observations on the duration.  The bottom red curve and vertical red lines indicate the location of telluric emission lines of OH and \nai.  The dashed magenta curves mark the expected location of any Doppler shifted D lines in the rest frame of Kepler-1520b, i.e. from any additional absorption due to \nai\ in a co-orbiting cloud.  The middle panel is the sum of the normalized spectra obtained during the transit, shifted into the rest-frame of the planet.  No such planet-related absorption lines are obvious in the data.     

\begin{figure}
	\includegraphics[width=\columnwidth]{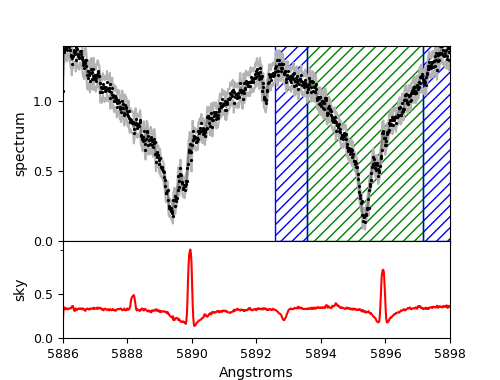}
    \caption{Same as Fig. \ref{fig:kepler-1520_sumspec} but for the transit of K2-22b on UT 26 January 2016.}
    \label{fig:k2-22-20160126_spec}
\end{figure}

\begin{figure}
	\includegraphics[width=\columnwidth]{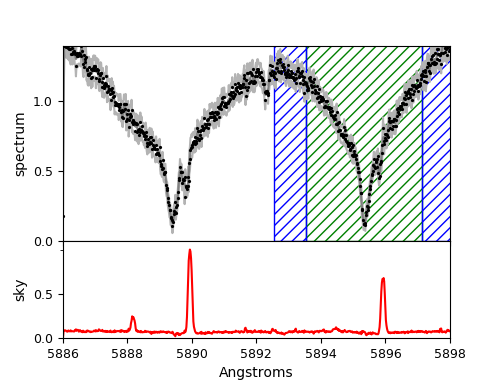}
    \caption{Same as Fig. \ref{fig:kepler-1520_sumspec} but for the transit of K2-22b on UT 29 January 2016.}
    \label{fig:k2-22-20160129_spec}
\end{figure}

\begin{figure}
	\includegraphics[width=\columnwidth]{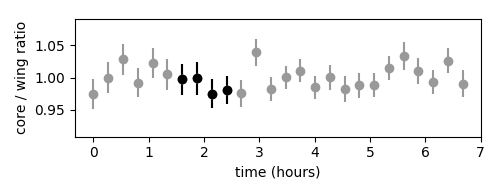}
    \caption{Same as Fig. \ref{fig:kepler-1520_time} but for the transit of K2-22b on UT 26 January 2016.}
    \label{fig:k2-22-20160126_time}
\end{figure}

\begin{figure}
	\includegraphics[width=\columnwidth]{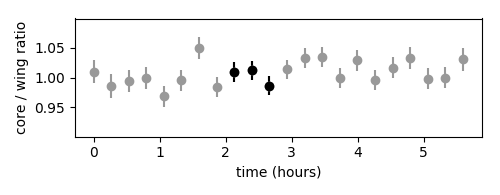}
    \caption{Same as Fig. \ref{fig:kepler-1520_time} but for the transit of K2-22b on UT 29 January 2016.}
    \label{fig:k2-22-20160129_time}
\end{figure}

\begin{figure}
	\includegraphics[width=\columnwidth]{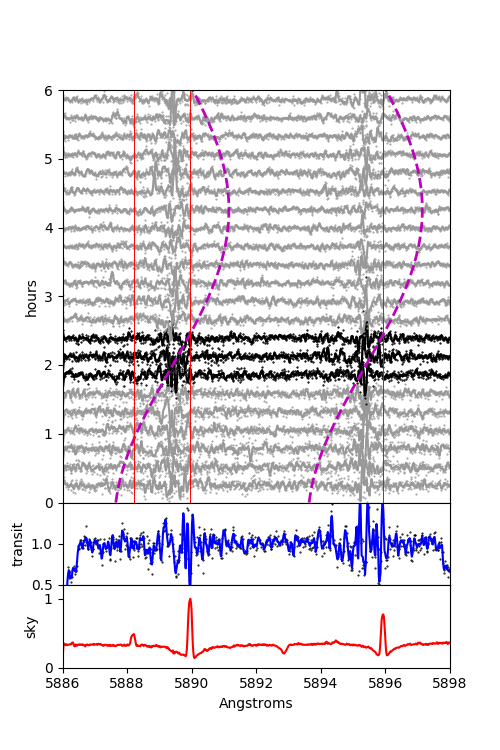}
    \caption{Same as Fig. \ref{fig:kepler-1520_series} but for the transit of K2-22b on UT 26 January 2016.}
    \label{fig:k2-22-20160126_series}
\end{figure}

\begin{figure}
	\includegraphics[width=\columnwidth]{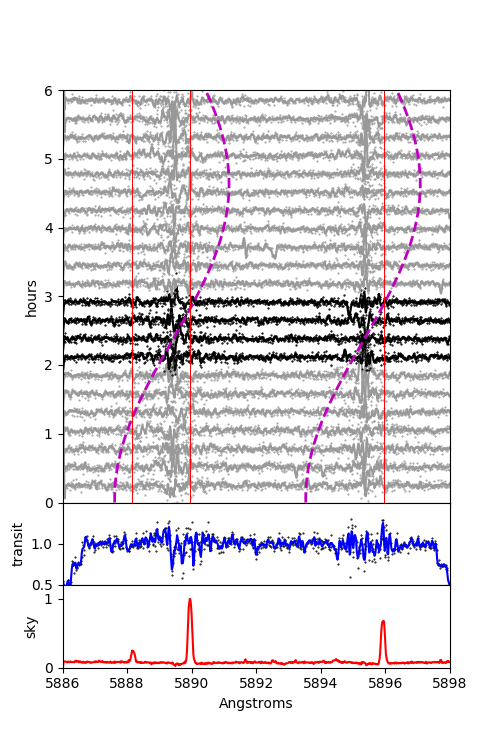}
    \caption{Same as Fig. \ref{fig:kepler-1520_series} but for the transit of K2-22b on UT 29 January 2016.}
    \label{fig:k2-22-20160129_series}
\end{figure}

Figures \ref{fig:k2-22-20160126_spec} and  \ref{fig:k2-22-20160129_spec} show the spectra of K2-22 obtained during the two transits of its evaporating planet on UT 26 and 29 January 2016.  Figures \ref{fig:k2-22-20160126_time} and  \ref{fig:k2-22-20160129_time} show the Na I D1 line flux ratio time series for those transits, and Figs. \ref{fig:k2-22-20160126_series} and  \ref{fig:k2-22-20160129_series} show the individual spectra.  In neither of the transits was a significant planet-associated signal observed.   

\begin{figure}
	\includegraphics[width=\columnwidth]{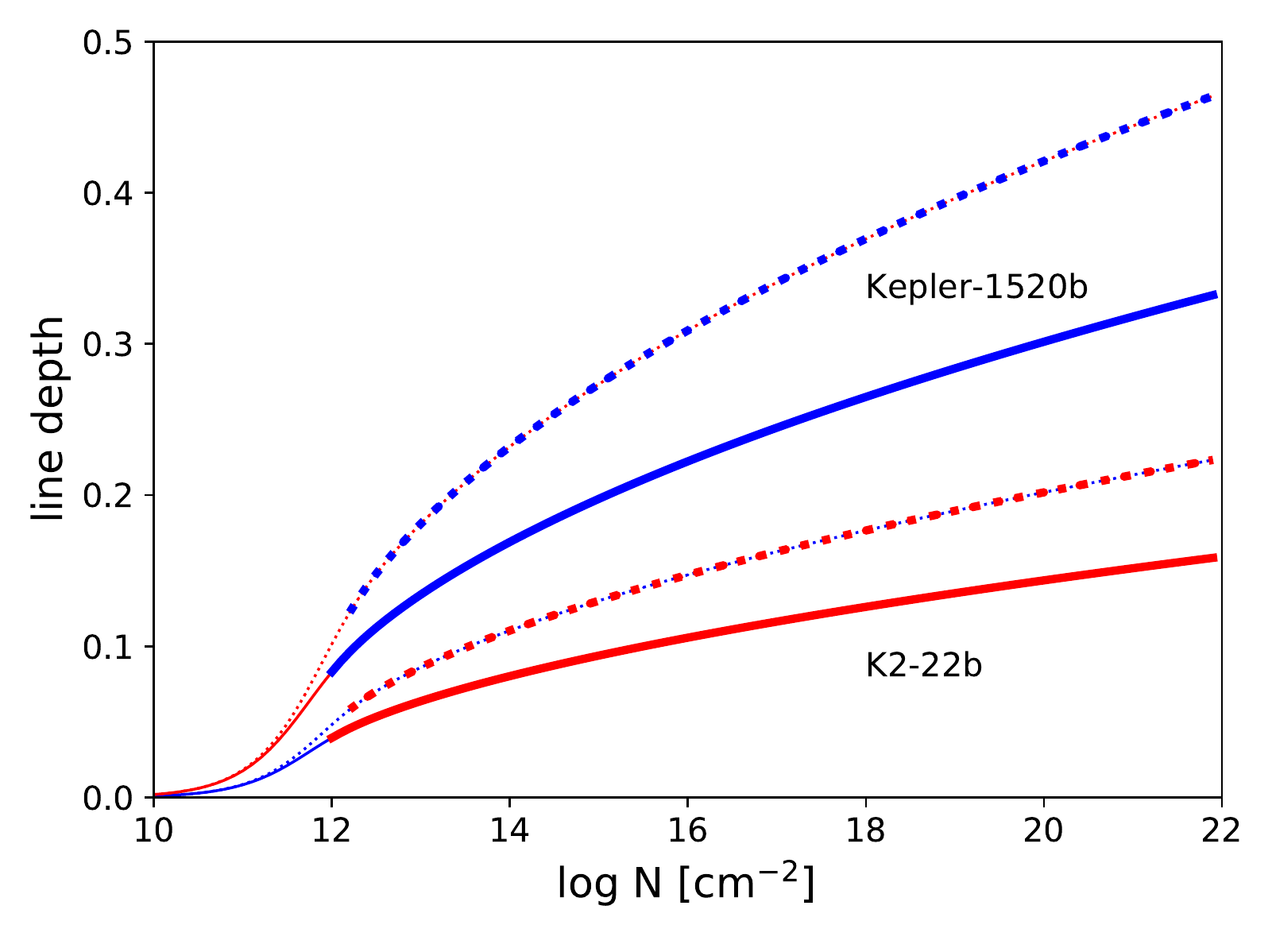}
    \caption{Predicted line depth of neutral sodium D lines in $R=45,000$ spectra with 900-sec integrations for a cloud co-moving with Kepler-1520b (upper pair of blue lines) and K2-22b (lower pair of red lines) that completely obscures the host star.  Doppler broadening at two temperatures (1000 K, solid lines, and 2000 K, dotted lines) are considered.  Line are thicker where the gas is optically thick ($\tau > 3$) at the line centre.  For partially obscuring clouds the line depth is multiplied by the obscuration fraction.  The equivalent width can be recovered by multiplying the line depth by 0.2\AA.}  
    \label{fig:cog}
\end{figure}

To determine the sensitivity of our observations we calculated the line depth produced by the transit of a co-orbiting cloud of \nai\ that completely obscures the star with a given column density $N$, and then injected lines with different depths into the actual spectra.   The line depth calculation (Fig. \ref{fig:cog}) assumes thermal broadening of the line at a temperature of 1050K, the 50\% condensation temperature into Na$_2$S \citep{Lodders2010}, an instrument resolution of 45,000, and Doppler broadening due to acceleration of the cloud over the 900 sec individual integrations (equivalent to degrading the resolution to 14000 for Kepler-22b and 7000 for K2-22b).  For $N < 10^{12}$~cm$^{-2}$ the lines are optically thin and the equivalent width and line depth are proportional to $N$ (Fig. \ref{fig:cog}).  The line depth is the equivalent width divided by the effective spectral resolution (including Doppler blurring); this is $\sim 10$ times wider than thermal broadening and when the lines become optically thick the line depth is $\sim 0.1$.   Hotter gas produces a deeper line because thermal broadening is greater, and gas associated with Kepler-1520b produces a deeper line than for K2-22b because the orbital period is longer and the Doppler acceleration over the integration time is less (Fig. \ref{fig:cog}).  Although the lines are saturated, line depth continues to (more slowly) increase at \nai\ column density $> 10^{12}$~cm$^{-2}$ because of substantial contribution by the wings.  If the cloud occults only a fraction of the stellar disk then the line depth is reduced proportionally.  

\begin{figure}
	\includegraphics[width=\columnwidth]{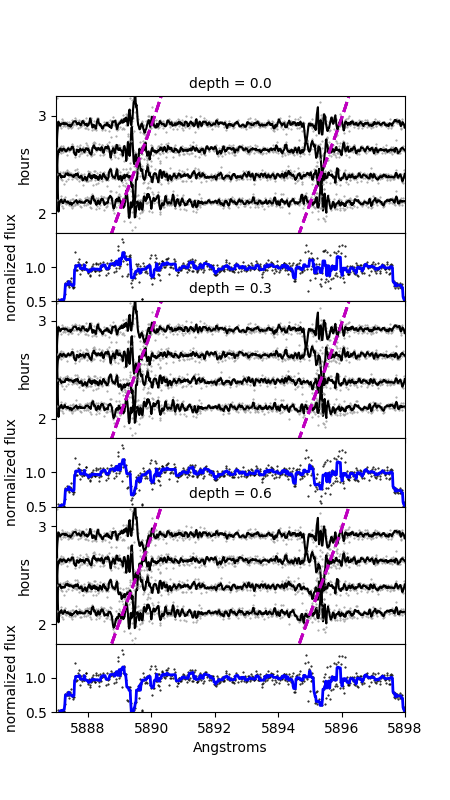}
    \caption{Simulations with synthetic \nai\ lines inserted into the actual spectra for Kepler-1520 with the expected Keplerian Doppler shift and thermal plus instrumental and Doppler broadening.  The top two panels show the original spectra plus the rest-frame shifted, summed spectra during transit, while the middle and bottom pairs of panels contain lines with depths of 0.3 and 0.6.}  
    \label{fig:detect}
\end{figure}

In Fig. \ref{fig:detect} we simulated a Na I sodium cloud co-moving with K2-22b by adding an unresolved line to the predicted location in the actual transit spectra, for the case of no line (top) and lines with depths of 0.3, and 0.6, accounting for the additional broadening of the line due to the finite integration time.  While a line with a depth of 0.6 should be readily identified in the spectra, a line with a depth of 0.3 would be a marginal detection.  Using numerical Monte Carlo experiments with a null (flat) spectrum plus the noise pattern from the spectra obtained of K2-22b during the second transit we find that a line with a depth of 0.4 could be validated (false positive rejection) with a confidence of 95\%.  Then using ``injection" experiments, we found that a line with a depth of 0.5 would be detected 94\% of the time.  Referring to Fig. \ref{fig:cog}, we would only expect to be able to detect a \nai\ column density $\gtrsim 10^{20}$~cm$^{-2}$ (in the most optimistic scenario, e.g., nearly complete stellar occultation).   

\section{Expectations for Sodium Around Evaporating Exoplanets}
\label{sec:expectations}

Here we estimate the column density of neutral \nai\ that might be observed in the vicinity of an evaporating rocky planet, provided instrument resolution and Doppler blurring are not factors.  On such a planet, high day-side temperatures produce a silicate vapor atmosphere that moves in a turbulent flow towards the terminator where partial re-condensation may occur \citep{Castan2011} but a significant fraction escapes to space.  This turbulent wind entrains and lofts dust grains that are accelerated from the planet by radiation pressure, forming a tail that causes the actual transit signal as it crosses our line of sight to the star.  Temperatures at the substellar point of the dark surface of a synchronously rotating, airless body at the orbits of Kepler-1520b and K2-22b are 2020K and 1900K, respectively, well above the $\approx$1500K solidus and 1670K viscosity transition in silicates, thus most evaporation will occur from a surface magma ocean \citep{Kite2016}.  

Sodium is a minor but significant component in primitive (undifferentiated, unmelted) chondritic meteorites \citep[0.5\%,][]{Lodders2010} and the primitive Earth mantle \citep[0.27\%,][]{Mcdonough1995}.  Sodium and the other alkali elements are highly incompatible elements that concentrate in silicate melts and hence can be expected to be overabundant, at least initially, in a magma ocean or crust (about 2.5\% for the case of Earth).  \citet{Schaefer2009} predicted that at 2000K, an atmosphere in equilibrium with a volatile-depleted mantle that otherwise resembles the bulk silicate Earth would be predominantly Na, with minor concentrations of molecular and atomic oxygen.  Evaporation of dust grains once they leave the surface and experience full illumination (equilibrium temperatures will exceed 1400K) and ultraviolet radiation \citep{Yakshinskiy2004} could also release gaseous Na.

There are two deviations from this simple scenario:  First, volatile elements such as sulfur or carbon might continue to make a significant contribution to a wind even from a depleted mantle, as demonstrated by continued sulfur outgassing on Io  \citep{Battaglia2014}.  Second, continued evaporation from the surface could deplete comparatively volatile elements and enrich in refractory elements.  \citet{Kite2016} described different regimes for exchange between the interior, magma ocean and atmosphere of a hot planet depending on the substellar temperature and FeO abundance in the mantle.  For $T<2400$K and FeO abundance in the range observed in the inner Solar System, evaporation of lighter, more volatile elements drives vertical sinking and mixing in the mantle ocean, but its lateral heat budget is unaffected by the wind.  In this regime, Na depletion of the magma ocean will proceed on a timescale \citep{Kite2016}
\begin{equation}
    \tau_d \sim \frac{d g R_p f_{\rm Na} \rho}{v P_{\rm eq}},
\end{equation}
where $d$ is the depth of the ocean, $g$ the surface gravity, $R_p$ the planet radius, $f_{\rm Na}$ the concentration of Na in the melt, $\rho$ the density of the melt, $v$ the wind speed, and $P_{\rm eq}$ the equilibrium pressure over the melt(a function of the concentration in the melt and temperature).  Since the pressure at the bottom of the magma ocean is $d g \rho$ and set by equality to the temperature excess $\Delta T = T_s - T_m$ divided by the rise in lock-up temperature vs. pressure ($\beta \sim 10^{-4}$~K~Pa$^{-1}$), then 
\begin{equation}
\tau_d \sim \frac{\Delta T R_p }{\beta v P_{\rm eq}}.
\end{equation}
Taking $\Delta T = 300$K, $v = 1$ \kms\ (sound speed for atomic Na gas), and $P_{\rm eq} = 10^{-3}$~Pa \citep{Schaefer2009}, the depletion time is $<1$~Myr.  On longer timescales the abundance of Na in the wind will be limited by the rate at which it is introduced into the magma ocean by melting to replace the loss by evaporation of the major constituents, e.g. SiO ($P_{\rm eq} \sim 10^{-6}$ Pa).  If this mass loss is set by the rate required to replenish the transiting dust in one orbital time then evaporation rates for Kepler-1520b and K2-22b are $> 2\times 10^{11}$~g~sec$^{-1}$ \citep{Perez-Becker2013,Sanchis-Ojeda2015}.  The concomitant rate of neutral sodium release for mantle-like material is then $f = 10^{31}$~sec$^{-1}$.

The column density of gaseous Na I that could produce any D-line absorption will be set by a balance between production, ionization by ultraviolet (UV) photons, and recombination.  Thermal ionization is unimportant at these temperatures.  Consider a spherically symmetric flow $f$ of neutral sodium atoms escaping  at velocity $v$ from the planet.  If the photoionization lifetime is $\tau_{\rm uv}$, then the ionization fraction $X$ follows:
\begin{equation}
\label{eqn:ionization}
    \frac{d}{dr}\left(r^2 n(r) v X\right) = r^2 n(r)\left[\frac{1-X}{\tau_{\rm uv}} - Rn(r)X^2 \right] 
\end{equation}
where $n(r)$ is the total space density of neutral and singly-ionized Na and $R$ is the recombination rate constant.  This assumes that the cloud is optically thin to ionizing photons, the only source of electrons is ionized sodium (as an abundant, readily ionized element) and that there are no other sinks (such as dust particles) of electrons.  Since $f = 4\pi r^2 n(r)v$, Eqn. \ref{eqn:ionization} can be re-written as:
\begin{equation}
    \label{eqn:ionization2}
    \frac{dX}{dr} = \frac{1}{v\tau_{\rm uv}}\left[1 - X - \frac{R \tau_{\rm uv} f}{4\pi r^2 v} X^2\right].
\end{equation}

\begin{figure}
	\includegraphics[width=\columnwidth]{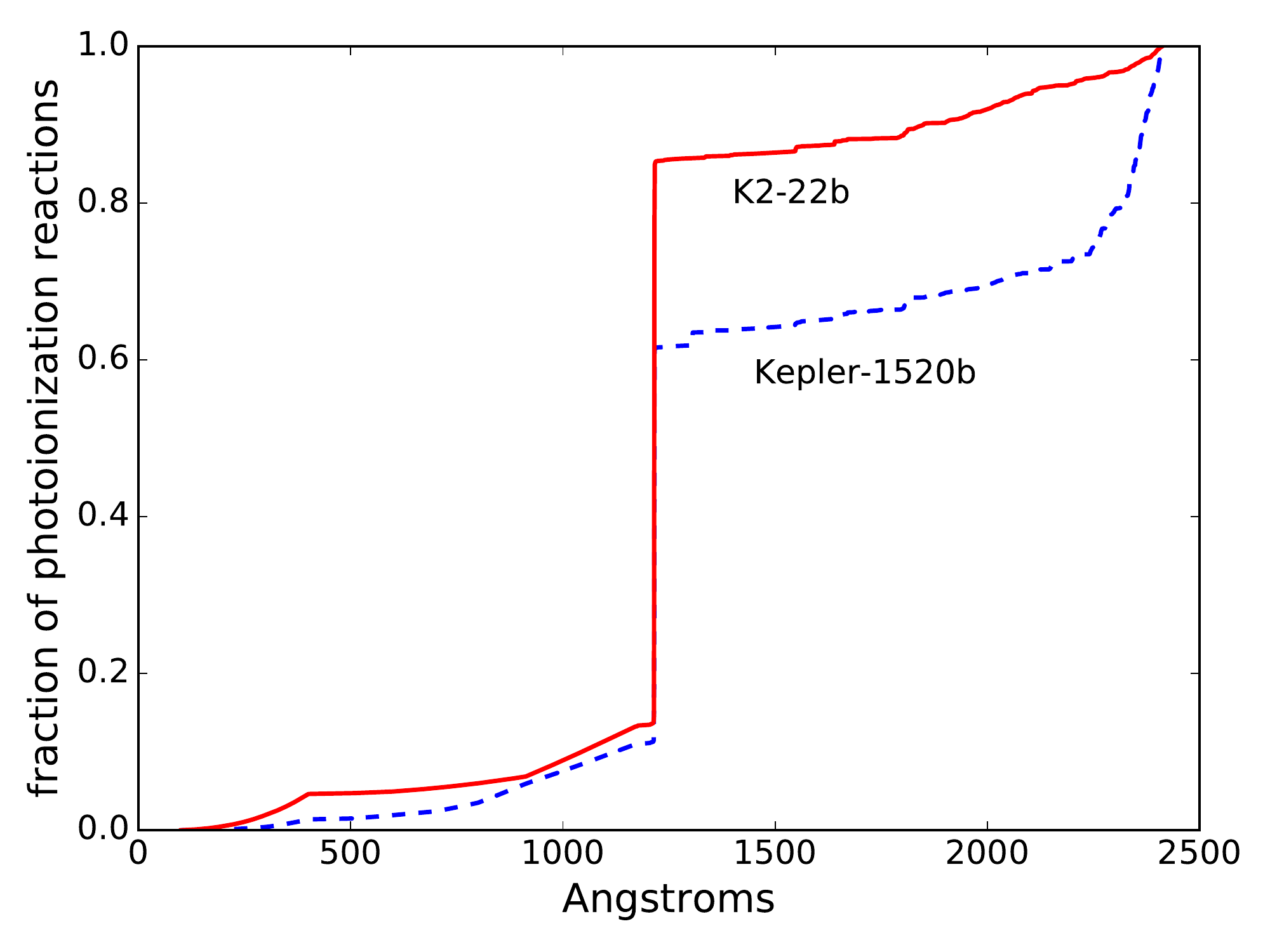}
        \caption{Cumulative distribution of photoionizations with wavelength for \nai\ atoms in the orbits of Kepler-1520b (dashed blue) and K2-22b (solid red), using the UV spectra of the proxy stars HD~85512 and GJ 667C, respectively.  The Lyman $\alpha$ line at 1214\AA\ is responsible for a significant fraction of the reactions.}
    \label{fig:ionize}
\end{figure}

Neutral sodium is ionized by UV photons with energies exceeding 5.14 eV ($\lambda < 2412$\AA).  To estimate the unshielded lifetime $\tau_{\rm UV}$ the stellar spectrum at higher energies (shorter wavelengths) must be convolved with the energy-dependent photoionization cross-section.  Kepler-1520 and K2-22 are too distant and faint to have been detected by the \galex\ UV mission \citep{Martin2003} or had UV spectra obtained with the STIS or COS instruments on \emph{HST}.  Instead, we used old, slowly-rotating stars of similar spectral type with UV spectra in the MUSCLES Treasury Survey \citep{France2016,Youngblood2016,Lloyd2016}.   The K6 dwarf HD 85512 was selected as the stand-in for Kepler-1520, and the M1.5 dwarf GJ 667C was chosen as a proxy for K2-22.  Stellar surface intensities were calculated by multiplying by the ratio of the proxy distance to stellar radius squared, and then the intensities at the orbits of Kepler-1520b and K2-22b were calculated by dividing by $(a/R_*)^2$.    These intensities were then convolved with the Na photoionization cross-section from \citet{Yeh1985} and \citet{Yeh1993} to calculate $\tau_{\rm UV}$:  1290 sec for Kepler-1520b, and 3930 sec for K2-22b.  About half of the ionization is provided by the Lyman-$\alpha$ line in the Kepler-1520 proxy and about two-thirds for the K2-22 proxy, and the majority of the remainder is from photons at wavelengths $\gtrsim 2000$\AA\ (Fig. \ref{fig:ionize}).  Since intrinsic Lyman $\alpha$ line strengths are difficult to measure and intrinsically variable with magnetic activity, these estimates are very approximate.  Using the recombination rate formulations of \citet{Badnell2006}, we estimate $R = 3.7 \times 10^{-9} \sqrt{T/1000}$~cm$^{3}$~sec$^{-1}$.   If the dispersal speed is comparable to the thermal speed then $v = \sqrt{T/1000}$~km~s$^{-1}$.  

\begin{figure}
	\includegraphics[width=\columnwidth]{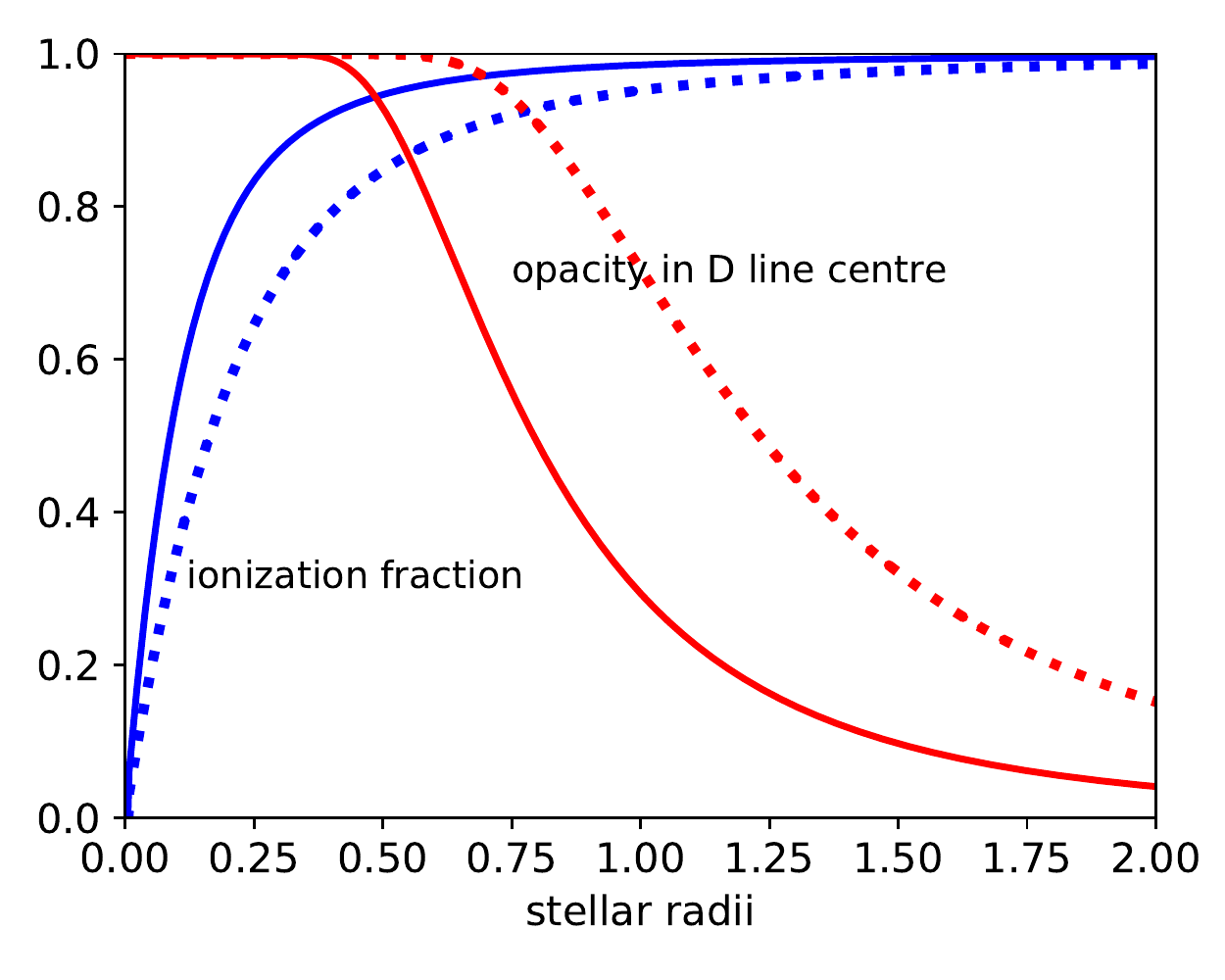}
        \caption{Ionization fraction (blue lines) of an expanding flow of sodium atoms ($1 \times 10^{31}$ sec$^{-1}$ released from Kepler-1520b (solid line) and K2-22b (dashed line) vs. distance in units of the host star radius.  Recombination rates assume that sodium atoms are the only source and sink for electrons.  The opacity (one minus transmission) in the D lines of Na I vs. the impact parameter of the line of sight from the source is also plotted as red lines.}
    \label{fig:ionize2}
\end{figure}

Figure \ref{fig:ionize2} shows the ionization fraction vs. radius (in units of the host star) for spherical clouds emanating from Kepler-1520b (solid line) and K2-22b (dashed line), using the values for $f$, $\tau$, $R$, and $v$ given above.   Also plotted is the opacity (one minus transmittance) in the \emph{resolved}, thermally-broadened D lines vs. impact parameter.  If the source is centred on the disk of the occulted star (mid-transit), the disk-averaged centre depth of a resolved line is 0.65 for Kepler-1520b and 0.92 for K2-22b, justifying the assumption of (nearly) complete occultation of the stars made in Sec. \ref{sec:results}.  The photoionization lifetime could be prolonged by self-shielding, if the sodium column is optically thick to ionizing photons, or scattering of UV photons by accompanying dust particles.  On the other hand, a higher velocity (e.g., due to acceleration of the partially ionized gas with the stellar wind) will shrink the zone of opaque Na I and possibly make it undetectable.  In our case, instrumental resolution (0.13 m\AA) and Doppler blurring of the lines (0.4-0.8 \AA) reduce these depths by up to an order of magnitude, but these results show that \nai\ could be detected with the appropriate instrument resolution and integration time.     

\section{Summary and Future Prospects}
\label{sec:discussion}

\begin{itemize}

\item We obtained high-resolution visible-wavelength spectra during transits of the ``evaporating" planets Kepler-1520b (one transit) and K2-22b (two transits) to search for absorption by neutral sodium associated with the dust that produces the transit-like signals seen in the broad-band filters of the \kepler\ telescope.  No significant differential absorption relative to the photosphere spectrum was detected. 

\item Our simulations show that we would have been able to readily detect a Doppler-shifted line with a depth $\ge 0.5$.  However, detection was limited by line dilution because of the lower resolution of the instrument relative to the thermal broadening of the lines at the expected temperature of $\ge$1000K, and blurring of the lines by the acceleration of the planet (and presumably, the cloud) during the finite exposure time, and would have been detected only in the most favorable of scenarios we considered.

\item Kepler-1520b and K2-22b are likely to be synchronously rotating and the highest temperatures, which control the surface pressure of the evaporative atmosphere, will be found at the substellar point.  Presuming that lighter volatiles have been depleted, the atmospheres are too thin to transport significant heat, substellar temperatures exceed 2000K, and evaporation occurs from a surface magma ocean.  Although Na is concentrated in silicate melts, the magma ocean would become depleted in $\sim$1~Myr and release of Na is limited by total mass loss and replacement by melting at the ocean bottom.

\item Assuming an Earth-like composition, a total mass loss rate of 1\,$M_{\oplus}$\,Gyr$^{-1}$, as suggested by the rate of dust production, corresponds to the release of $10^{31}$~sec$^{-1}$ atoms of Na.  A model of a spherically symmetric wind of Na with UV ionization and recombination predicts that the cloud of neutral Na will extend a significant fraction of the stellar radius.  The depth of the D lines averaged over the stellar disk is 0.6-0.9, but the finite instrument resolution and integration times dilute the lines by up to an order of magnitude and to an undetectable level.

\item The abundance of \nai\ could be lower if the winds from these planets are intermittent \citep[not necessarily the same as intermittent dust production,][]{Schlawin2018}, or Na is greatly depleted in the silicate mantles of Kepler-1520b and K2-22b due to formation close to the parent star.   Other effects that should be considered are ongoing degassing of volatiles from the mantles, shielding of Na from UV by dust particles, and acceleration and shaping of the Na cloud by the stellar wind.

\end{itemize}

Echelle spectra permit the search for transient absorption signals among other lines.  These include the resonant lines of neutral potassium K~I of 7665 and 7699\AA.  Since Na and K have similar condensation temperatures they are reasonably expected to be co-occurring in a gaseous plume from a disintegrating planet.  While the reach of HDS spectra includes these lines, the effects of fringing at these redder wavelengths prevented useful analysis of our data. 

Spectroscopic investigation of `disintegrating' planets is primarily limited by the faintness of the host stars and the briefness of the transit, limiting the available signal-to-noise.  The \emph{TESS} mission \citep{Ricker2014}, presently carrying out a monitoring survey of nearly the entire sky for $>$27 day intervals, could detect similar objects around brighter stars.  \emph{TESS} will observe many subgiant stars \citep{Stassun2018} with evolving luminosities that could drive evaporation and eventual destruction of any close-in small planets.  During its prime mission, \tess\ will observe Kepler-1520 (but not K2-22, which is close to the ecliptic plane), however the 1-hour photometric error for this $V=16.7$ star will be $>2$\%, larger than the transit depth of Kepler-1520b ($\lesssim 1$\%).    Three planned Extremely Large Telescopes (the European Extremely Large Telescope, the Thirty Meter Telescope, and the Giant Magellan Telescope) should begin observations in the next decade and will have collecting power that exceed Subaru by at least an order of magnitude.

Kepler-1520b and KOI-2700 are among about 200,000 stars monitored by \kepler\ during its prime mission, and among about 32,000 fellow K dwarfs \citep{Berger2018}: thus the rate among K dwarfs, at least, is $6 \times 10^{-5}$.  Since the duration of the phenomenon is $\sim$100~Myr \citep{Perez-Becker2013} or about 2\% of the typical age of the star, this suggests that $\sim$0.3\% of such stars have planets smaller than Mercury on $\lesssim$1~day.  This would suggest that the radius distribution for planets is roughly flat in log radius below 0.8\rearth\ \citep[c.f.,][]{Sanchis-Ojeda2014}.   

We estimated the occurrence of evaporating planets on transiting orbits among \tess\ targets.  We assumed that at a mass loss above that of the lower bound for \kepler-1520b (0.1 $M_{\oplus}$ Gyr$^{-1}$) the dusty tail, if not the planet, would be detectable.  We use a Jean's criterion for massive hydrodynamic escape of silicate vapor and dust from the planet \citep{Volkov2011}, 
\begin{equation}
    \lambda \equiv \frac{G M_p \mu}{R_p k_B T} < 2.8
\end{equation}
where $G$ is the gravitational constant, $M_p$ and $R_p$ are the planet mass and radius, $\mu$ the mean molecular weight of the gas, and $T$ the temperature of the atmosphere.  We adopt $\mu = 30$ (nucleon units), appropriate for a mix of SiO and MgO gas \citep{Perez-Becker2013}.  At $\lambda > 2.8$ escape is limited by the Maxwellian distribution of the gas, and is neglected.  Since $R_p$ decreases more slowly with $M_p$ than linear, at a given $T$ hydrodynamic escape occurs below a certain value of $M_p$ (roughly Mercury-Mars size in the case of \kepler-1520b).  During hydrodynamic escape ($\lambda < 2.8$) the mass loss rate is taken to be:
\begin{equation}
    \dot{M} \sim \rho c_s R_p^2,
\end{equation}
where $\rho$ is the atmosphere density and $c_s$ is the sound speed in the atmosphere, computed with ratio of specific heats $\gamma = 1.3$.  The density is $\rho = p \mu/(k_B T)$, with the pressure:
\begin{equation*}
    p = {\rm exp} \left(b - \frac{\mu_s L}{k_B T}\right),
\end{equation*}
where $\mu_s \approx 169$ is the mean molecular weight in the crust that corresponds to the dominant molecules in the atmosphere.  The criterion for detection is that current mass loss has to be above the detection threshold ($\dot{M} \gtrsim 0.1 M_{\oplus}$ Gyr$^{-1}$) but \emph{cumulative} mass loss over the star's history (excluding the pre-main sequence phase) did not destroy the planet: i.e., that the planet's required original mass was not above the largest size of rocky cores.  We set this threshold to 1.5\rearth: there are planets larger than this but these are larger because they possess H-He envelopes \citep{Weiss2014,Rogers2015}, and planets in the intermediate range are relatively uncommon \citep{Fulton2017}.  

\begin{figure}
	\includegraphics[width=\columnwidth]{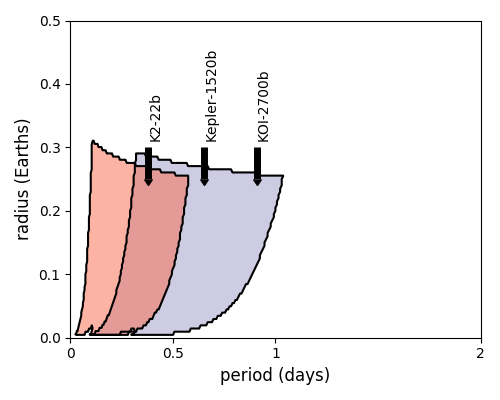}
        \caption{Estimated zones where an evaporating rocky planet of a given radius and orbital period could be detected, i.e. it is evaporating at least 0.1$M_{\oplus}$ Gyr$^{-1}$ but its past evaporation has not led to its destruction.  The purple zone is for a K dwarfs like those hosting the evaporating planets \kepler-1520b and KOI-2700b, and the red zone is for the M dwarf host of K2-22b.  Arrows indicate the orbital periods of the three planets; the radii are not known but are thought to be smaller than Mercury (0.36\rearth).}
    \label{fig:evaporate}
\end{figure}

Effective temperatures and luminosities are taken from the \tess\ Candidate Target List (CTL) list v. 7.02 \citep{Stassun2018}\footnote{CTL v7 does yet incorporate \gaia\ DR2 parallaxes}.  These are compared to the main sequence and post-main sequence tracks (age $>10^8$ yr) from the Dartmouth Stellar Evolution Program \citep{DOtter2008} using a Gaussian prior for the metallicity distribution with mean of -0.2 dex and standard deviation 0.25 dex \citep{Casagrande2011}.  The best-fit mass track and current luminosity are combined with a given orbital period to calculate the stellar irradiance and the planet's surface temperature, assuming negligible albedo (i.e. forward scattering into the dusty atmosphere dominates back-scattering out of it) and inefficient heat re-distribution (i.e., we assume the formation of the atmosphere is driven by conditions at the substellar point).  The mass evolution of the planet is back-calculated using the best-fit stellar evolution track to 100~Myr.  These calculations are performed for a range of orbital periods and current planet radii.  Figure \ref{fig:evaporate} shows the case for the three known disintegrating planets.  Above the upper boundary the planet is too massive for significant evaporation.  To the right the planet is too distant and cool.  To the left evaporation is too rapid and no such planet should exist at the present epoch.  The ranges where sub-Mercury size planets are predicting to be evaporating at the present epoch correctly correspond to the location of the known objects. 

With that success, we assume the cumulative planet distribution with $P<10$~days scales as $P^{0.77}$ based on the distribution of Earth-size planets compiled by \citet{Sanchis-Ojeda2015} and uniform distribution with radius.  We then average over a randomly selected sample of 10,000 stars from the CTL, and arrive at a mean probability of $7 \times 10^{-5}$, which is essentially the \kepler\ result.  Thus among 200,000 stars that \tess\ might observe with 2-minute cadence we would expect $\sim$14 detections.  This does not account for the limited photometric precision of \tess\ relative to the signal produced by the dust clouds.  \tess\ will also survey a much larger number of stars at 30 min cadence of its Full Frame Images, which include many subgiant stars.  It is among these systems that additional evaporating planets might be found and their tenuous atmospheres explored by the next generation of ground-based telescopes.

\section*{Acknowledgements}

The initial phase of the project were undertaken while EG was a visiting scientist supported by the Swiss National Science Foundation at the Observatoire de Versoix at the University of Geneva.  Additional support was provided by NASA grant NNX11AC33G (Origins of Solar Systems) to EG.  This work was supported by Japan Society for Promotion of Science (JSPS) KAKENHI Grant Number JP16K17660. We thank Kevin France for help with UV spectra from the MUSCLES  Treasury.

%%%%%%%%%%%%%%%%%%%%%%%%%%%%%%%%%%%%%%%%%%%%%%%%%%



%\bibliographystyle{mnras}
%\bibliography{references_master} % if your bibtex file is called example.bib

% Don't change these lines
\bsp	% typesetting comment
\label{lastpage}
\end{document}